\documentclass[graybox,natbib,nosecnum]{svmult}
\bibpunct{(}{)}{;}{a}{}{,} 

\usepackage{amsmath}
\usepackage{amssymb}
\usepackage{upgreek}
\usepackage{wrapfig}

\pdfoutput=1   

\usepackage{mathptmx}       
\usepackage{helvet}         
\usepackage{courier}        
\usepackage{type1cm}        

\usepackage{makeidx}         
\usepackage{graphicx}        
\usepackage{multicol}        
\usepackage[bottom]{footmisc}
\usepackage[normalem]{ulem}	
\usepackage{hyperref}  
\definecolor{Red}{rgb}{0.75,0,0}
\definecolor{Blue}{rgb}{0,0,0.75}
\hypersetup{colorlinks=true,linkcolor=Blue,citecolor=Blue,
            filecolor=Blue,urlcolor=Blue}

\usepackage{soul}   

\usepackage{color}

\makeindex             

\begin{document}

\title*{Disintegrating Rocky Exoplanets}
\author{Rik van Lieshout \& Saul Rappaport}
\institute{Rik van Lieshout \at Institute of Astronomy, University of Cambridge, Madingley Road, Cambridge CB3 0HA, UK \\ \email{lieshout@ast.cam.ac.uk}
\and Saul Rappaport \at MIT Kalvi Institute for Astrophysics and Space Research and Physics Department \\ \email{sar@mit.edu}}

%
%
\maketitle

~

\abstract{We discuss a new class of exoplanets that appear to be emitting a tail of dusty effluents.  These disintegrating planets are found close to their host stars and have very hot, and likely molten, surfaces.  The properties of the dust should provide a direct probe of the constituent material of these rocky bodies.}

\section{A New Class of Exoplanets--Overview}
\label{sec:overview}

Three exoplanets have been discovered with the {\em Kepler} mission that are inferred to have tails of dusty effluents trailing behind (or even ahead) of them in orbit about their host star \citep{2012ApJ...752....1R,2014ApJ...784...40R,2015ApJ...812..112S}.  These objects, known respectively as {KIC~12557548b} (KIC~1255b for short), {\mbox{KOI-2700b}}, and {\mbox{K2-22b}}, have several characteristics in common.  The orbital periods are all quite short at 15.7, 22, and 9.5 hours, respectively.  The depths of the occultations as they pass in front of their host stars range from 0.02\% to 0.5\%, which imply blocking areas of 1$-$18 times that of the Earth.  However, the occultations are asymmetric about their midpoint, and are otherwise inconsistent with transits of solid-body planets.  Importantly, the transit depths of all three objects vary with time -- KIC~1255b and \mbox{K2-22b} exhibit rapid variations in depth from transit to transit, while those of \mbox{KOI-2700b} vary slowly with time over the four years of the main {\em Kepler} mission.  

All of these observational characteristics point toward the occulter being an elongated {tail of dusty material} emanating from an underlying hot rocky exoplanet.  However, it should be noted up front that at present it is difficult to ascertain either the mass or the radius of these {`disintegrating' planets}, as they have come to be known.  The inferred mass-loss rates from the planets, based on the amount of dust required to yield such significant extinction of the host star, is in the range of $10^{10 \pm 1}$ g s$^{-1}$.  This translates to $\sim$4 lunar masses per Gyr.  In turn, this implies that for rocky bodies of radius $25$, $250$, and $2500$ km, the disintegration lifetimes would be $10^3$ years, 1~Myr, and 1~Gyr, respectively.  The extremely short implied lifetimes for bodies under 1000 km, compared to the lifetimes of their host stars, suggest a rather small likelihood for discovering such objects.  Therefore, in this work we assume that the disintegrating exoplanets are at least the size of Ceres, a minor planet in our solar system with a mass of 1\% that of the Moon, and a radius of $\sim$500 km.  

The recently discovered transits of white dwarf \mbox{WD~1145+017}, which show some of the same properties as the disintegrating exoplanets, are discussed in ``Transiting Disintegrating Planetary Debris around WD 1145+017'' (Vanderburg and Rappaport; this handbook).

\section{Observed Transit Profiles and Variability}
\label{sec:profiles}

\begin{figure}
\centering
\includegraphics[scale=.75]{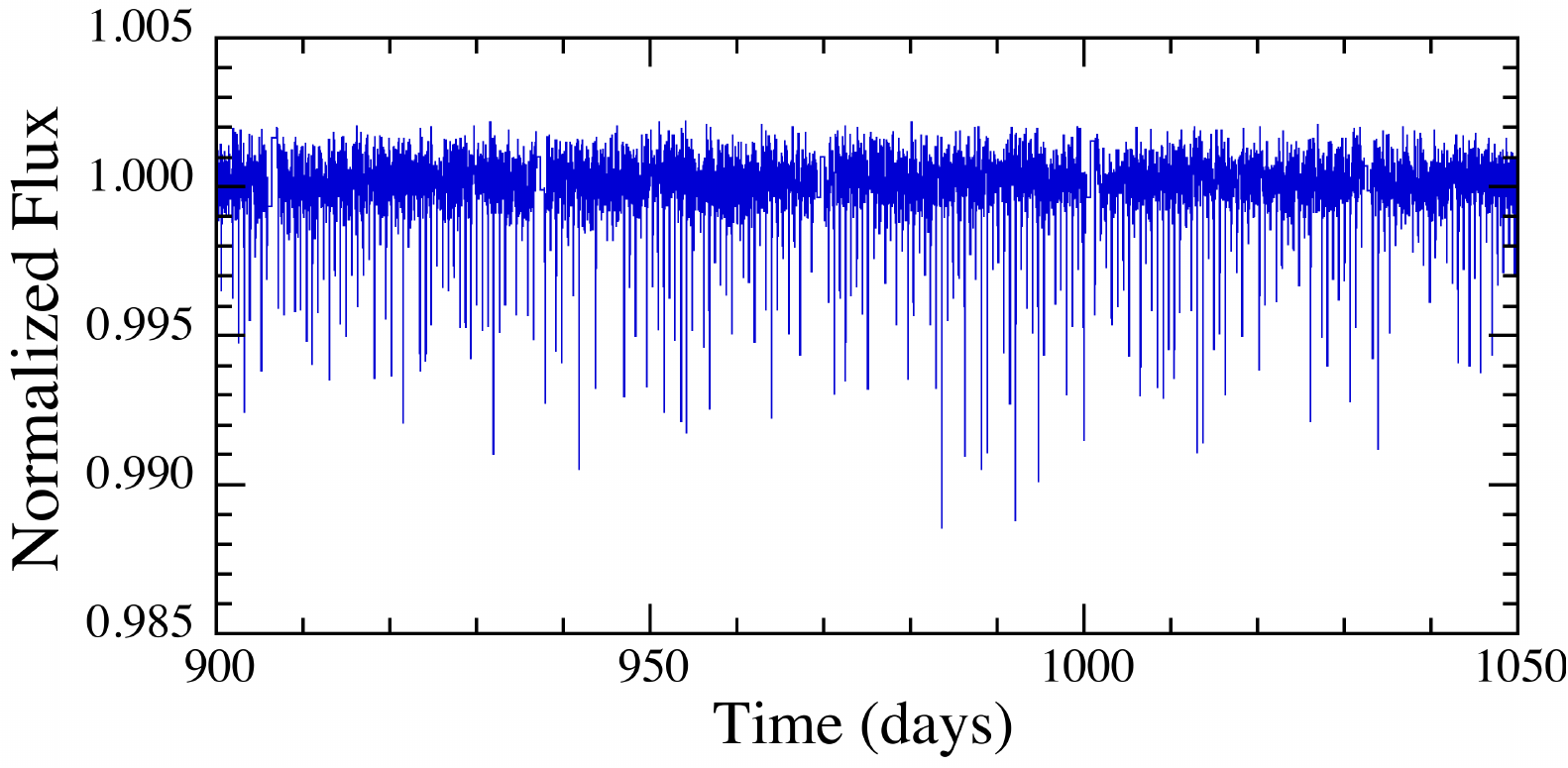}
\caption{Flux lightcurve for KIC~12557548 spanning 150~days of the {\em Kepler} mission. The 22-day starspot rotation signature has been filtered out.  Note the highly variable transit depths.  }
\label{fig:lc}
\end{figure}

KIC~1255b and \mbox{K2-22b} were discovered in the {\em Kepler} and K2 data as a direct result of the extreme {variability of their transit depths} \citep{2012ApJ...752....1R,2015ApJ...812..112S}.  These variations are often dramatic, even from transit to transit.
Figure~\ref{fig:lc} shows a 150-day portion of the light curve of KIC~12557548. Over the entire four years of the {\em Kepler} mission, there were
more than 2000 transits by KIC~1255b, with depths ranging between 0 and 1.2\%, as well as several week-long intervals when there are no detectable transits.

\begin{figure}
\centering
\includegraphics[scale=0.505]{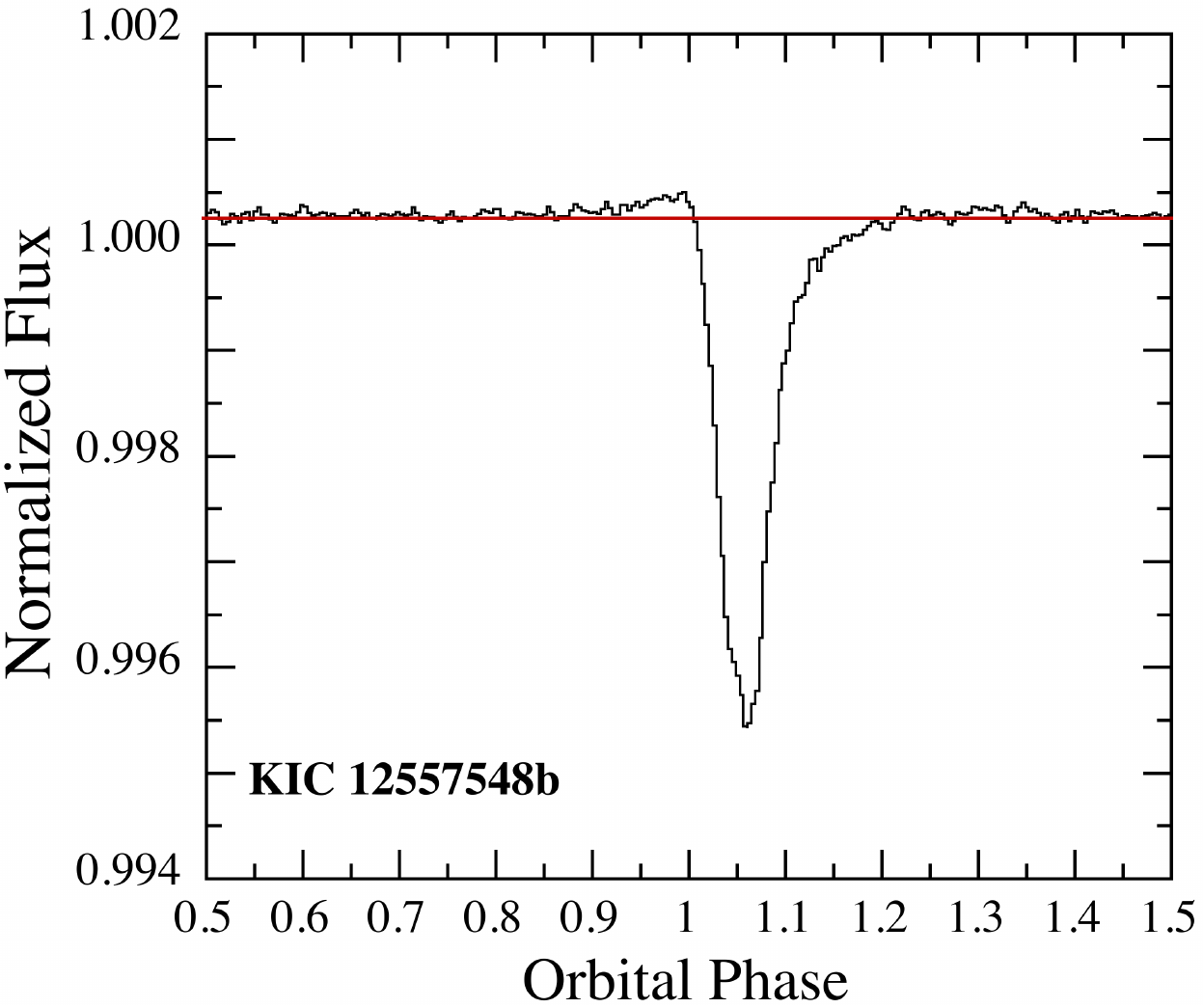} \vglue0.01cm \hspace{0.12cm} \includegraphics[scale=0.505]{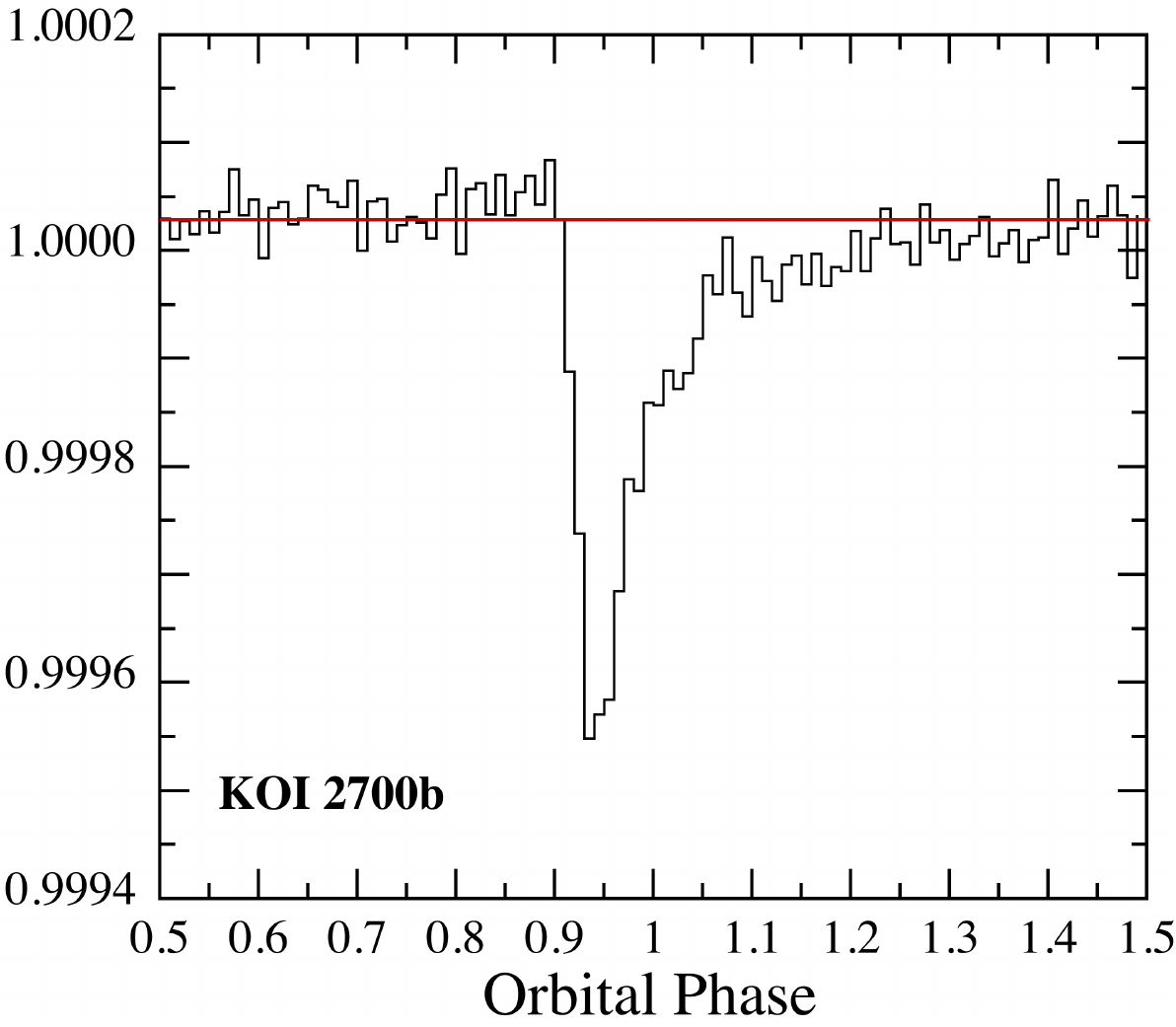} \vglue0.01cm \hspace{0.2cm} \includegraphics[scale=0.505]{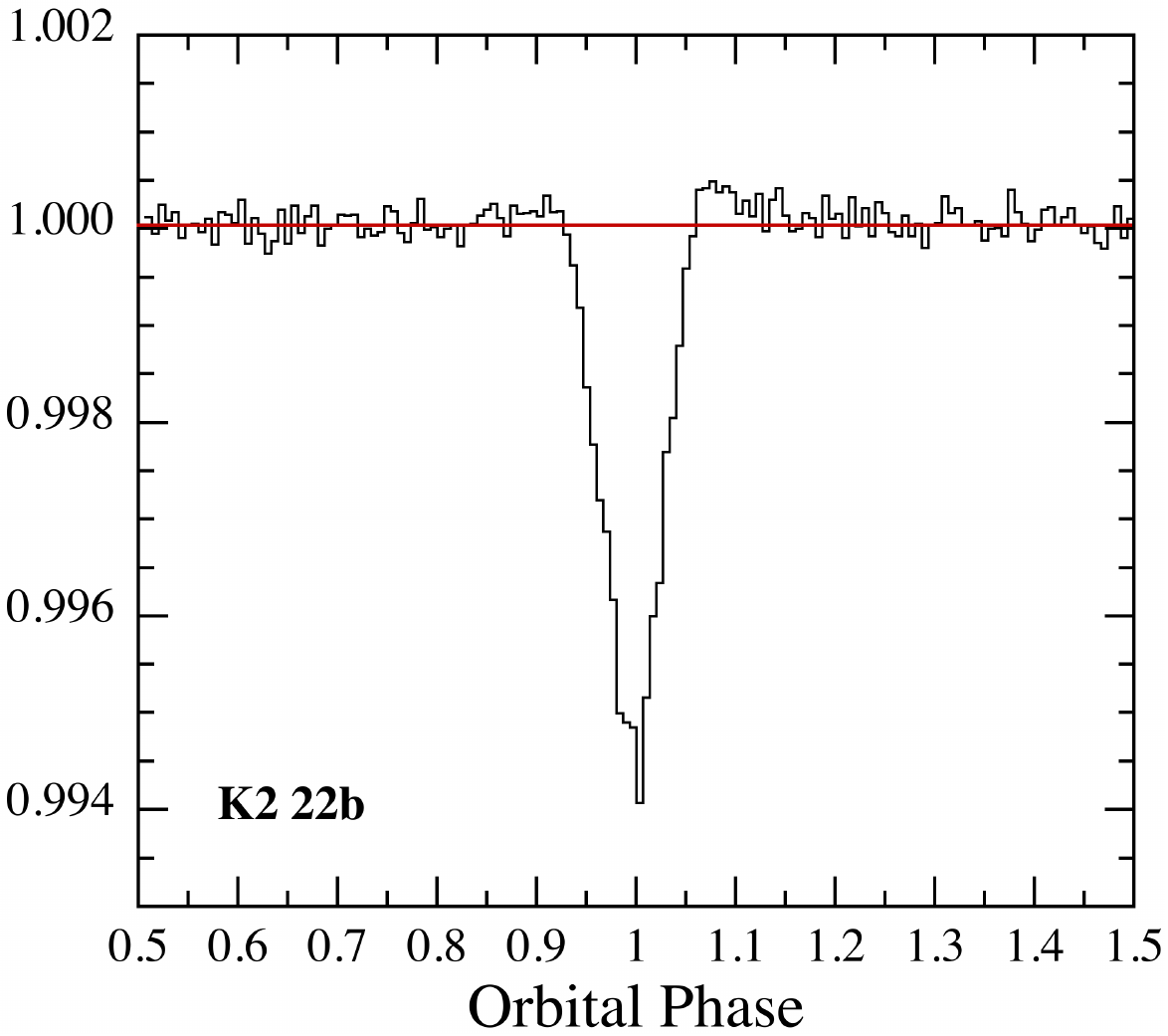}
\caption{Mean transit profiles for KIC~1255b, \mbox{KOI-2700b}, and \mbox{K2-22b}.  The first two are characterised by a sharp ingress followed by a long egress.  Note that the mean transit depth of \mbox{KOI-2700b} is about an order of magnitude more shallow than that of the other two objects.  Also note the small pre-ingress `bump' for KIC~1255b and the post-egress `bump' for \mbox{K2-22b}.}
\label{fig:profiles}
\end{figure}

When all of the data are folded modulo the period of the transits, we find the mean transit profile shown in the top panel of Fig.~\ref{fig:profiles}.  The transit profiles for \mbox{KOI-2700b} and \mbox{K2-22b} are shown for comparison in the middle and bottom panels of the same figure.  The data for all these transits were recorded in the {\em Kepler} long-cadence mode (with an integration time of $\sim$1/2 hr).  Therefore, the transit widths, which range from $\sim$1.5 to 4 hours, are somewhat distorted by the finite integration time.  Nonetheless, for KIC~1255b and \mbox{KOI-2700b} the transit profile has a distinctly sharper ingress and a substantially longer {egress tail}.  The statistics on KIC~1255b are sufficiently good that one can also discern a distinct bump before the ingress.  We explain this feature in the following sections.  The transit profile for \mbox{K2-22b} is more symmetric; however, there is a statistically significant bump just after the transit egress.  We provide an explanation for this feature as well in the following sections.

For future reference, we summarise some of the known properties of these dusty-tailed planets and their host stars in Table \ref{tbl:dusty}.  For all three objects, the host star is of mid-K to early-M spectral type with relatively low luminosity.  As mentioned, the orbital periods of these planets are all short with $P_{\rm orb}$ ranging from 9$-$22 hours.  Importantly, the value of $a_{\rm p}/R_*$, which is the ratio of the orbital distance of the planet to the radius of the host star, ranges from 3.3 to 5.9.  Another way of stating this is that the orbits are only 1$-$2.5 times the star's diameter above the surface of the host star.

\begin{table}
\begin{center}
\caption{Comparative Properties of Dusty Planets and Their Host Stars}
\begin{tabular}{lcccc}
\hline
\hline
Parameter & Symbol [units] & KIC~1255b & \mbox{KOI-2700b} & \mbox{K2-22b} \\
\hline
\hline
\multicolumn{5}{l}{Host star parameters} \\
\hline
Stellar temperature & $T_{\rm eff,\star}$ [K] & $4550 \pm 135$ & $4300 \pm 140$  & $3830 \pm 100$ \\
Surface gravity & $\log g$ [cgs] & $4.62 \pm 0.04$ & $ 4.71 \pm 0.05 $  & $ 4.65 \pm 0.12 $ \\
Metallicity & [Fe/H] & $-0.2 \pm 0.3$ & $-0.7 \pm 0.3$ & $0.03 \pm 0.08$ \\
Stellar mass & $M_\star$ [M$_\odot$] &  $0.67 \pm 0.06$ & $0.55 \pm 0.04$ & $0.60 \pm 0.07$ \\
Stellar radius & $R_\star$ [R$_\odot$] & $0.67 \pm 0.06$ & $0.54 \pm 0.05$ & $0.57 \pm 0.06$ \\
Stellar luminosity & $L_\star$ [L$_\odot$] & $0.17 \pm 0.04$ & $0.09 \pm 0.02$  & $0.063 \pm 0.008$ \\
Stellar rotation period & $P_{\rm rot}$ [days] & 22.9 & 11.0 & 15.3 \\
Maximum $\beta$ ratio of dust$^{\rm a}$ & $\beta_{\rm max}$ & 0.19 & 0.12 & 0.07 \\
Radial velocity variations$^{\rm b}$ & $K$ [m/s] & $< 150 $ & ... & $< 280$ \\
\hline
\multicolumn{5}{l}{Planet: light curve properties} \\ 
\hline
Orbital period & $P_{\rm orb}$ [hr] & 15.68 & 21.84 & 9.146 \\
Transit depth (range) & $\delta$ [\%] & $0$--$1.4$  & $0.031$--$0.053$ & $0$--$1.3$ \\
Mean transit depth & $ \left< \delta \right> $ [\%] & 0.5 & 0.036 & 0.55 \\
Variability & &  fast  & slow  & fast \\
Long egress & &  yes &  yes & no \\
Pre-ingress bump & &  yes &  ? & weak  \\
Post-egress bump &  &  no &  no & yes \\
\hline
\multicolumn{5}{l}{Planet: derived parameters} \\ 
\hline
Planet radius & $R_{\rm p}$ [R$_\oplus$] & $\lesssim 1$ & $< 0.5$ & $< 3$ \\
Semi-major axis & $a_{\rm p}$ [AU] & 0.0129(4) & 0.0150(4) & 0.0088(8) \\
Scaled semi-major axis & $a_{\rm p}/R_\star$  &  $4.3 \pm 0.4$ &  $5.9 \pm 0.4$ & $3.3 \pm 0.2$ \\
Transit impact parameter & $b$ & $0.6 \pm 0.1$ & $<$1.0 & 0.42$-$0.78 \\
Angular radius of star$^{\rm c}$ & $\theta_\star$ [$^\circ$] &  13 &  10 & 17 \\
Tail length$^{\rm d}$ & $\theta_{\rm tail}$ [$^\circ$] & $\sim$$10$--$15$  & $\sim$24 & $< 6$  \\
Planet's peak temperature$^{\rm e}$ & $T_{\rm eff,p}$ [K]  &  2100 &  1850  & 2100 \\
Planet's dust mass-loss rate & $\dot M_{\rm d}$ [$10^{11}$ g/s] & $\sim$2 & $\sim$0.06 & $\sim$2 \\
\hline
\end{tabular}
\label{tbl:dusty}
\end{center}
{Notes:} Data were compiled from \citet{2012ApJ...752....1R}; \citet{2012A&A...545L...5B}; \citet{2013A&A...557A..72B}; \citet{2014A&A...561A...3V} for KIC~1255b; from \citet{2014ApJ...784...40R} for \mbox{KOI-2700b}; from \citet{2015ApJ...812..112S} for \mbox{K2-22b}; and from \citet{2014ApJS..211....2H} for most stellar parameters. \\
(a)~Where $\beta$ is the radiation-pressure-to-gravity force ratio. \\
(b)~Based on the scatter in the RV measurements of \citet{2014ApJ...786..100C,2015ApJ...812..112S}. \\
(c)~Angle subtended by the stellar radius as seen from the planet. \\
(d)~Angular cross-section density $e$-folding angle, assuming exponential decay. \\
(e)~Equilibrium temperature at the planet's substellar point, $T_{\rm eff,p} \equiv T_{\rm eff,\star} \sqrt{R_\star/a_{\rm p}}$.
\end{table}

\section{Conclusion That These are Dust Tails}
\label{sec:tails}

Shortly after KIC~1255b was discovered, it was concluded that by far the most plausible explanation for the observed properties of the transits is that they are due to a dust tail trailing behind the planet in its orbit.  Considering also the transit profiles of the other two objects, \mbox{KOI-2700b} and \mbox{K2-22b}, we can list five main reasons for believing that the transits in these objects are due to {dusty tails}.  These include: (1) strictly periodic transits whose depths vary with time; (2) transit shapes that are distinctly asymmetric with a long egress tail (in two of the objects); (3) positive `bumps' in flux either prior to, or following, the transit that can be explained via forward scattering by dust (two of the objects); (4) no evidence for any close binary companions; and (5) no evidence for any secondary eclipses or occultations.

No known close binary star system with periods $\lesssim$ a day, has properties anything like those enumerated above.  Moreover, no hard-body planet transits resemble these properties either.  Dust provides a natural explanation for temporal variations in the transit depths, asymmetric profiles with a tendency for long egress tails, and the small positive-going bumps in the light curves.  We return to discuss some of these explanations in more detail in the upcoming sections.  

Even though dust seems like a natural and convenient explanation for the phenomena we see in these transiting systems, this does not constitute a proof that dust is at the heart of the correct scenario.  However, we have spent considerable effort in trying to conjure up other explanations for the temporally varying {asymmetric transits}---but without success.  Nonetheless, it is wise to include consideration of other scenarios, even as evidence mounts for the dusty tailed planets.

\section{Explanation of Transit Profiles}

\begin{figure}
\centering
\includegraphics[scale=.95]{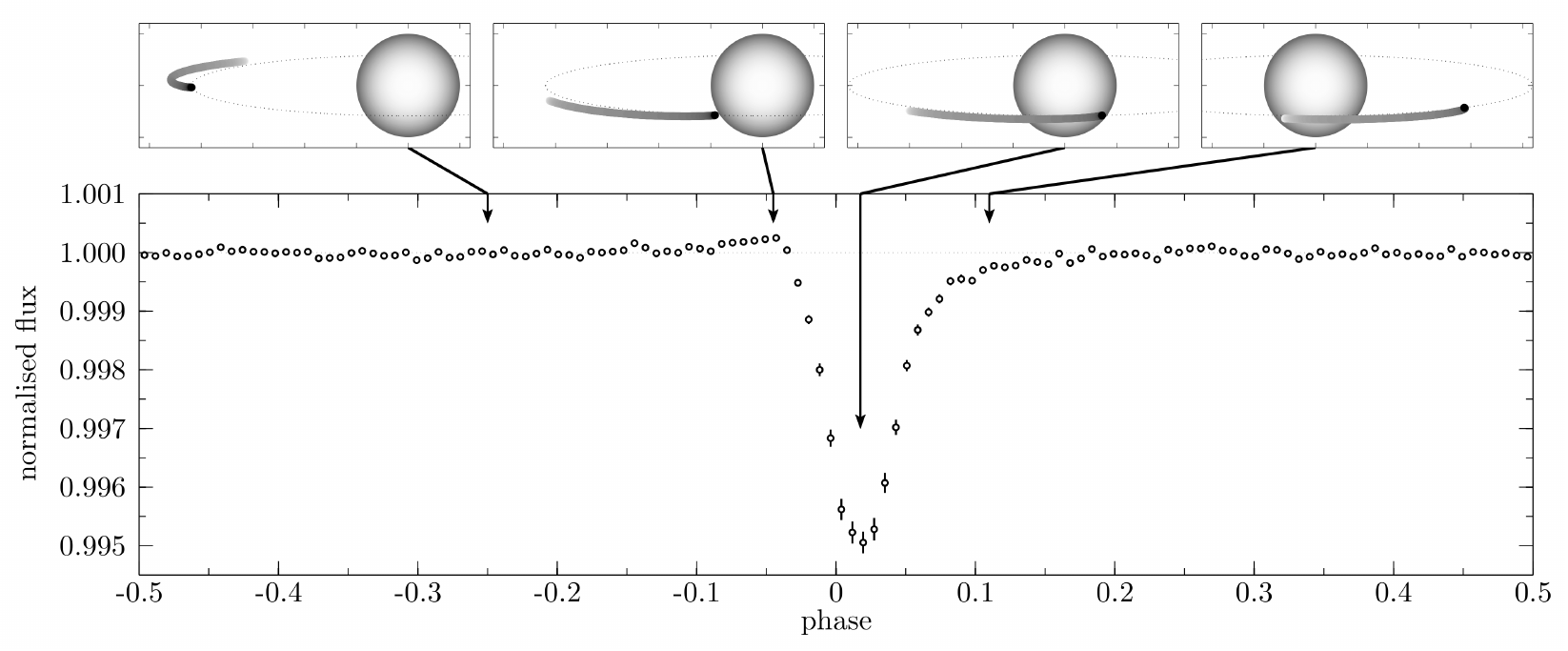}
\caption{Dust-tail geometry shown for a sequence of four orbital phases of the underlying planet \citep{2016A&A...596A..32V}.}
\label{fig:sketch}
\end{figure}

The long egress of two of the transit profiles (see Fig.~\ref{fig:profiles}) is explained by the fact that the densest part of the dust tail is presumably located near the hard body of the planet (the source of the dust), and this density falls off with angular distance behind the planet.  Therefore, once the planet crosses the limb of the host star the transit ingress starts relatively abruptly, and the transit depth continues to increase until the hard-body planet moves off the opposite limb of the star.  Following that, there is a slow recovery from the transit as the remainder of the dust tail moves off the disk of the star.  This sequence of events can be visualised in Fig.~\ref{fig:sketch}.

It is important to note that the optical depth of the dust tail decreases with distance behind the planet largely due to the effects of: (1)~sublimation of the dust grains; (2)~acceleration of dust grains away from the planet due to radiation pressure forces; and (3)~differing degrees of shearing due to radiation pressure forces exerted on dust grains of different sizes.  Because of conservation of angular momentum, the dust tail does not spread much in the direction perpendicular to the orbital plane.  However, as discussed in the next section, the tail does spread both in the radial and azimuthal directions due to radiation pressure forces by differing amounts for particles of different sizes.  Thus, it is inevitable that the optical depth of the dust tail will decrease with distance from the planet as the number density and grain sizes decrease due to dilution from shearing and by sublimation, respectively. 

The small, but very significant, {pre-transit bump} in KIC~1255b (Fig.~\ref{fig:profiles}), can be explained as follows.  Just prior to the ingress when the planet hard-body crosses the stellar limb, there is actually very little attenuation of the stellar flux.  However, some of the starlight that is then heading away from the Earth, but toward the dusty tail, can be scattered back into our line of sight.  We can expect such `{forward scattering}' to produce a `bump' on the light curve that is of order a few to 10\% of the transit depth \citep[see][]{2012ApJ...752....1R,2013A&A...557A..72B,2016MNRAS.461.2453D}. This forward scattering persists all the way through the transit, but that positive-going flux is dwarfed by the light extinction during the transit itself \citep{2013A&A...557A..72B,2015ApJ...812..112S,2016MNRAS.461.2453D}; therefore, one sees the effects of forward scattering only before and after the main transit event. 

The small bump after the egress of the transit in \mbox{K2-22b} (bottom panel in Fig.~\ref{fig:profiles}) is hypothesised to occur from a dust tail that extends predominantly in the opposite direction, i.e., a tail that {\em leads} the hard-body planet in its orbit.  This geometric configuration can come about if the dust particles are sufficiently small (or sufficiently large) that radiation pressure forces are negligible and the material emanating from the planet is preferentially directed toward the host star \citep{2015ApJ...812..112S}.  This turns out to be a reasonable assumption for the case where a Parker wind (see section ``Physics of the Planetary Outflow'') is responsible for driving matter away from the planet \citep{2012ApJ...752....1R,2013MNRAS.433.2294P}, or even in the case where free streaming metal vapours emanate from the planet's hemisphere that is facing the host star (see following sections).

\section{Dynamics of the Dust Tails}

In order to use the observed transit profiles of the dust tails to infer properties of the dust grains and the planets that release them, it is critical to understand the behaviour of the dust. In the following, we discuss the forces and processes that govern the evolution of the dust grains after their release from the parent body.

Dust grains are believed to be carried away from the planet by a gaseous outflow \citep{2012ApJ...752....1R,2013MNRAS.433.2294P}. Initially, the dust particles are entrained in the gas \cite[see Sect.~4.2 of][]{2013MNRAS.433.2294P}, but as the gas expands into a larger volume and becomes more tenuous the drag forces weaken, causing the dust to decouple from the gas. The motion of the gas with respect to the planet at the time of the decoupling determines the velocities with which the dust grains are launched into space. Since the planetary wind is driven by stellar heating, which is confined to the planet's star-facing hemisphere, the wind may be directed primarily toward the star. The magnitude of the launch velocity is poorly constrained, but should exceed the local escape speed of the planet.

Close to the planet, the planet's gravity will affect the dynamics of the dust. This is roughly confined to spatial scales of the planet's Roche lobe. Assuming a planet mass of 0.02~M$_\oplus$ \citep[an upper limit estimated for KIC~1255b by][]{2013MNRAS.433.2294P}, the Roche lobe extends to about 1~R$_\oplus$ from the planet. This is considerably smaller than the length of the dust tails, which in the case of KIC~1255b extends to more than 100~R$_\oplus$ behind the planet.

\begin{figure}
\centering
\includegraphics[scale=.35]{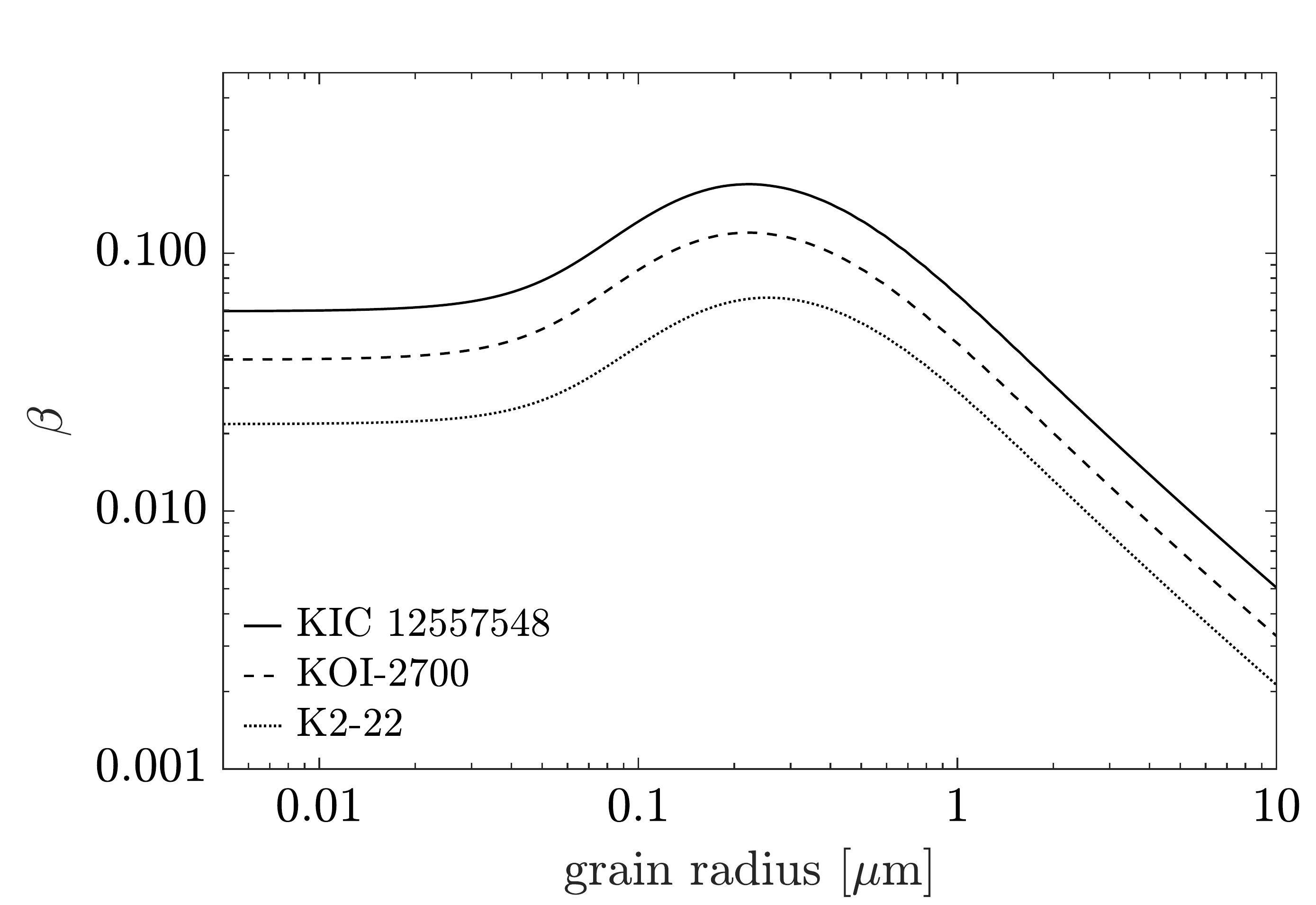}
\caption{Radiation-pressure-to-gravity force ratio $\beta$ as a function of dust grain radius for the three systems.}
\label{fig:beta}       
\end{figure}

Once the dust grains have left the immediate vicinity of the planet, their dynamics are dominated by stellar gravity and {radiation pressure}. The relative importance of these two forces is parameterised by $\beta$, the ratio between the norms of the direct radiation pressure force and the gravitational force. Since the radiation pressure force has the same scaling with distance as gravity ($ \propto r^{-2} $), $\beta$ is independent of position, and depends only on stellar parameters, grain size and the optical properties of the dust material (see, e.g., \citealt{1979Icar...40....1B}; Sect.~4.2 of \citealt{2014ApJ...784...40R}; Sect.~2.2 of \citealt{2014A&A...572A..76V}; Sect.~9.2 of \citealt{2015ApJ...812..112S}). Figure~\ref{fig:beta} shows how $\beta$ changes as a function of grain size, and how it differs among the three systems because of differences in stellar mass and luminosity.

The force of the {\em {stellar wind}} on the dust grains acts in the same way as radiation pressure and, in analogy to the radiation pressure $\beta$, it can be parameterised by the stellar-wind-to-gravity force ratio $\beta_{\rm w}$.  \citet[their Appendix~A.1]{2014ApJ...784...40R} estimate for \mbox{KOI-2700b} that the stellar wind is one or two orders of magnitude weaker than radiation pressure, although this is relatively uncertain. One of the reasons for the weakness of the stellar wind pressure is that at the radial distance of the disintegrating planet, the stellar wind is still accelerating and has not yet reached a very high velocity \citep[see Sect.~4.5 of][]{2013MNRAS.433.2294P}.

If launch velocities are negligible compared to the effects of radiation pressure, the orbit of a dust grain after being liberated from the planet is determined purely by its $\beta$ ratio.  Radiation pressure on dust grains has the same effect as a change in stellar mass. A grain that is released from the planet will initially have the same velocity as the planet, that is, the Keplerian velocity corresponding to the true mass of the star (assuming that the planets have circular orbits). When radiation pressure is taken into account, however, this velocity corresponds to the pericentre velocity of a larger, eccentric orbit around the star \citep[see Sect.~4.1 and Appendix~B of][]{2014ApJ...784...40R}.

Since the new orbit has a slightly longer orbital period than that of the planet, the dust will gradually lag behind. In the frame corotating with the planet, the grains move away from the planet, forming a tail behind it \citep[see also Fig.~1 of][]{2014A&A...572A..76V}.  Just after release, when the dust velocity is almost the same as that of the planet, the relative velocity will still be small, giving a larger angular density of dust grains close to the planet.

The orbit-averaged speed of a dust grain with respect to the planet due to radiation pressure is proportional to its $\beta$ ratio. Comparing this speed with the initial launch speed reveals whether the orbit of a dust grain is set primarily by radiation pressure or by the launching mechanism \citep[see Eq.~(23) of][]{2014A&A...572A..76V}. This, in turn, determines the shape of the dust cloud and hence that of the transit profile. Differences in stellar parameters can therefore explain the differences between the transit profiles of \mbox{K2-22b} and the other two objects.

\begin{figure}
\centering
\includegraphics[scale=0.55]{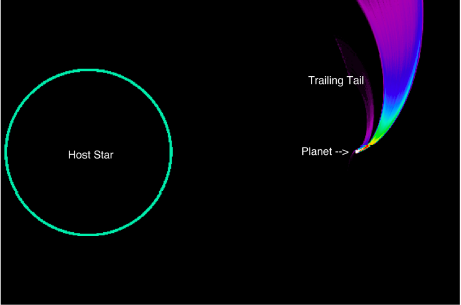} \vglue0.1cm \includegraphics[scale=0.55]{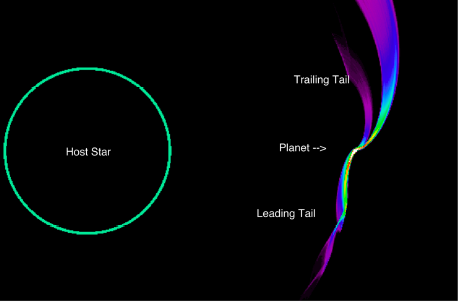}
\caption{Dust-tail simulations \citep{2015ApJ...812..112S}, as viewed from far above the planet's orbital plane. The top panel results from dust particles that feel significant radiation pressure forces, while for the bottom panel we arbitrarily multiplied each calculated value of $\beta$ by a factor of 1/4. The colour coding is proportional to the logarithm of the dust particle density with white-red the largest to blue-purple the lowest. The dust tails are shown in the rest frame of the orbiting planet (implicitly moving downward in the image).}
\label{fig:sims}
\end{figure}

To help better visualise the geometry and dynamics of the dust tails, we show dust-particle simulations in Fig.~\ref{fig:sims}. 
Particles with a range of sizes are launched roughly in the direction of the host star. They are then subject to radiation pressure and gravity, and their size is set to gradually decay to simulate the effect of sublimation (see Sect.~9.2 of \citealt{2015ApJ...812..112S} for more details of the simulations).
To show how the relative importance of radiation pressure governs the dust-cloud morphology, the $\beta$ ratios in one of the two simulations were scaled down by a factor four. In the top panel, with high $\beta$ ratios, radiation pressure dominates the dynamics, giving only a trailing dust tail. In the bottom panel, where radiation pressure is much less significant, there is both a leading and a trailing dust tail.

Finally, we note that the dust tails are usually assumed to be optically thin, or at most marginally optically thick, to the stellar radiation. The effects of possible self-shielding of the dust within the cloud on its dynamics are still under study. Likewise, the possibility and consequences of mutual collisions amongst dust grains are not yet studied in detail, but these were briefly discussed by \citet[their Sect.~9.5]{2015ApJ...812..112S}.

\section{Effects of Dust Sublimation}

Because of their proximity to the host star, the dust grains heat up to temperatures that are high enough to sublimate the material they are made of. {Sublimation} converts solid dust into gas, and hence causes the dust grains to gradually become smaller. As dust grains becomes smaller, they block less starlight, which is one of the reasons for the gradual egress in the light curves.
The rate at which the {extinction cross-section} goes down with decreasing grain size depends on the grain-size-to-radiation-wavelength ratio (see Sect.~5.1.1 of \citealt{2014ApJ...786..100C} for an overview of the different regimes).
Eventually sublimation destroys the dust grains, and this gives the tails their finite length.

As a dust grain becomes smaller, its $\beta$ ratio changes, influencing the dynamics. For sufficiently large particles, $\beta$ increases with decreasing grain radius $s$, following a power law ($\beta \propto s^{-1}$; see Fig.~\ref{fig:beta}). In this regime, sublimating grains will accelerate away from the planet \citep[see Eq.~(9) of][]{2014A&A...572A..76V}. The resultant gradually-decreasing angular number density of dust particles contributes to the gradual egress of the transit profile.
For smaller grains, $\beta$ no longer follows the same power law, leading to more complex dynamical effects.

\begin{figure}
\centering
\includegraphics[scale=.40]{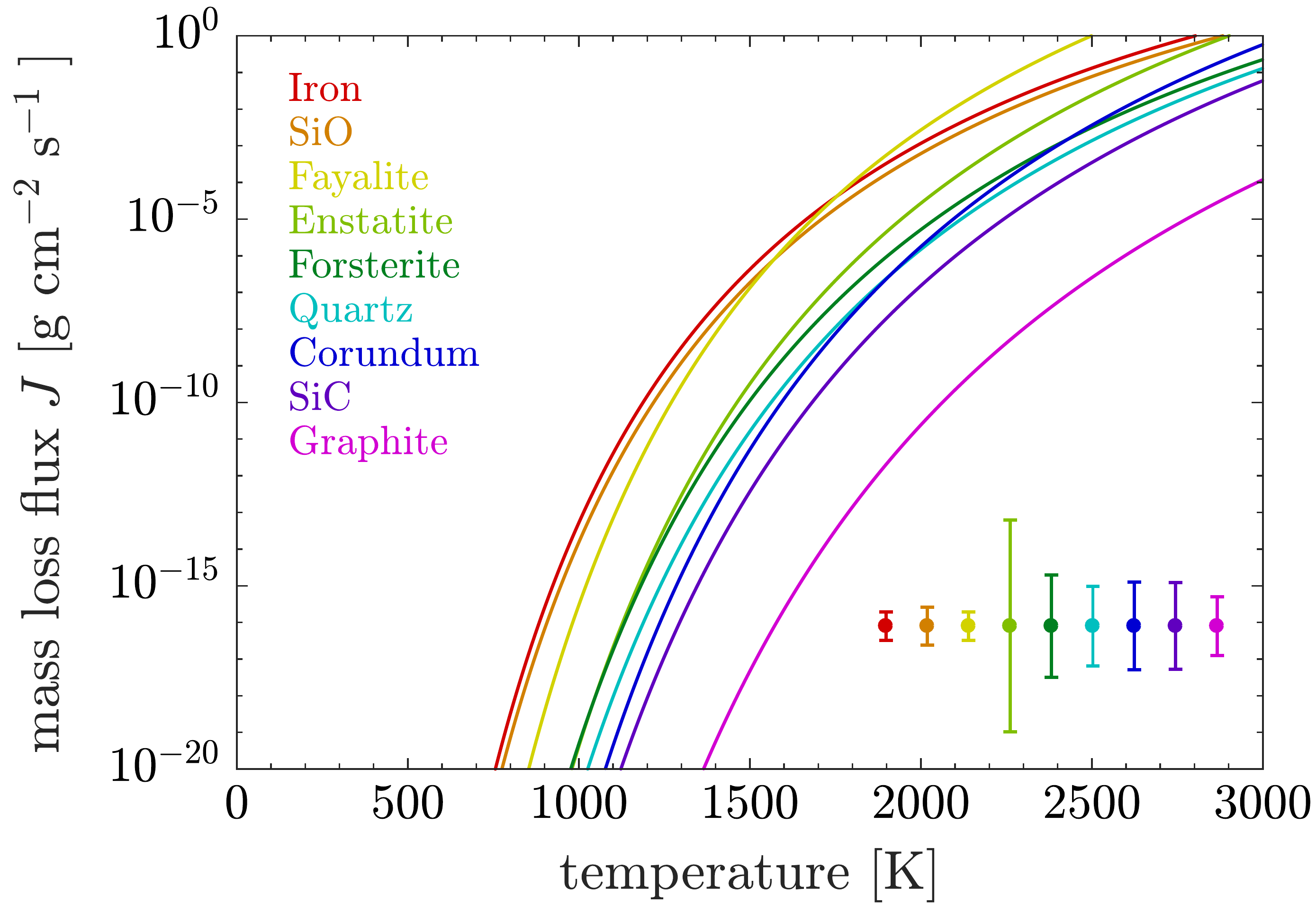}
\caption{Mass-loss fluxes vs. the grain equilibrium temperature during the sublimation process for different minerals. The error bars indicate typical uncertainties on the mass-loss flux, estimated at a temperature of 2000\,K by propagating the uncertainties on the underlying sublimation parameters.}
\label{fig:subl_rate}       
\end{figure}

The rate of sublimation depends very sensitively on a grain's temperature
(see Fig.~\ref{fig:subl_rate}),
which varies with distance to the star. Since dust grains follow eccentric orbits, their sublimation rate can vary significantly over their orbit \citep[see Fig.~2 of][]{2016A&A...596A..32V}. Dust grains are released at their periastron. If they survive the initial bit of their orbit, most of their orbit is spent somewhat farther away from the star, where the sublimation rate may be significantly lower. The grains only experience the highest sublimation rates again after one orbit when they return to periastron. This may be an explanation for the lifetime of dust particles in the tail of KIC~1255b, which is estimated to be close to one orbital period (see Sect.~4.3 of \citealt{2013MNRAS.433.2294P}; Sect.~3.1.1 of \citealt{2014A&A...561A...3V}).

{Dust temperature} also depends on the absorption and emission efficiencies of the grains. These, in turn, are determined by the grain size and the optical properties of the material that the grain is made of (i.e., its complex index of refraction). The absorption and emission efficiencies drop rapidly at wavelengths longer than a particle's circumference. Hence, small grains often cannot cool efficiently through infrared emission, while still being heated nominally by short-wavelength stellar radiation. As a result, they can reach equilibrium temperatures much higher than that of larger grains, which act more like black bodies. In the case of KIC~1255b, for example, 0.1\,$\upmu$m grains can reach temperatures close to 2000~K, while a 1\,$\upmu$m grain may have a temperature of around 1500~K \citep[this also depends strongly on composition; see Fig.~B.2 of][]{2016A&A...596A..32V}. The higher equilibrium temperature of small grains, together with the extremely high sublimation rate at these higher temperatures, may act as a sharp cut-off in the {grain size distribution}.

The heating of dust grains due to impact of {stellar wind} particles, as well as the {latent heat of sublimation}, are thought to be insignificant \citep[see Appendices A.2 and C.1 of][]{2014ApJ...784...40R}.
Optical depth effects may influence dust temperatures close to the planet.

\section{Inferences about Dust from the Transit Profiles}

In transits of disintegrating rocky exoplanets, the blocking cross-section is dominated by dust
(rather than the hard-body planet itself) and some of the properties of this dust can be inferred from the detailed shape of the {transit profile} and its wavelength dependence. While the spatial distribution of the dust is not known a priori, it can be either prescribed by an ad-hoc (but physically informed) morphological model or calculated from first principles.  By comparing synthetic transit light curves generated from such models to an observed one, it is possible to put constraints on {dust properties}, which, in turn, provide clues about the parent body.

To generate a {synthetic light curve} from a dust cloud model, one needs to consider the effects of both absorption and scattering of starlight by dust grains. In the simplest case of an extinction-only, one-dimensional dust-cloud model, the synthetic light curve is obtained by convolving the azimuthal distribution of extinction cross-section with the stellar emission profile at a given impact parameter. Possible sophistications that can be added to this basis model include (1)~the effect of ``forward'' scattering of starlight (i.e., into the line of sight), necessary to explain the bumps in the light curves; (2)~the imprint on the light curve of the hard-body planet; and (3)~the vertical extent of the dust cloud (over a range of impact parameters).
Table~\ref{tbl:lc_mdl} gives an overview of the different {light curve models} and their properties.

\begin{table}
\begin{center}
\caption{Summary of Light Curve Modelling Efforts}
\begin{tabular}{l|cccc|c|c}
\hline
\hline
 & \multicolumn{4}{c|}{KIC~1255b} & \mbox{KOI-2700b} & \mbox{K2-22b} \\
Model reference & B12 & B13 & V14 & V16 & R14 & S15 \\
\hline
Extinction & \checkmark & \checkmark & \checkmark & \checkmark & \checkmark & \checkmark \\
Forward scattering & \checkmark & \checkmark & & \checkmark & & \checkmark \\
Hard-body planet & & & & & \checkmark & \\
Vertical extent & & \checkmark & \checkmark & & & \\
Azimuthal distribution & E & E/P & E & T & E & E2 \\
\hline
\end{tabular}
\label{tbl:lc_mdl}
\end{center}
{Model references:
B12:~\citet{2012A&A...545L...5B}, see also Sects.~4.1 and 4.3 of \citet{2014A&A...561A...3V};
B13:~\citet{2013A&A...557A..72B};
R14:~\citet[their Sects.~4.4 and 4.5]{2014ApJ...784...40R};
S15:~\citet[their Sect.~9.3]{2015ApJ...812..112S};
V14:~\citet{2014A&A...561A...3V};
V16:~\citet{2016A&A...596A..32V}.
Azimuthal distribution prescriptions: E:~Exponential decay; E2:~Exponential decay, both leading and trailing the planet; P:~Power-law decay; T:~Tracking the dust-grain evolution by integrating the equations of motion and sublimation.}
\end{table}

The transit profile is determined primarily by the azimuthal distribution of extinction cross-section. Using the knowledge of dust dynamics and sublimation from the previous sections, this distribution can, under a number of assumptions, be derived.
In the regime of small, Rayleigh-scattering particles, it is slightly steeper than exponential decay \citep[Appendix~C.2 of][]{2014ApJ...784...40R}. In the large-particle regime, it is somewhat more shallow than exponential decay \citep{2014A&A...572A..76V}.
These findings support the use of the exponential-decay distribution assumed by most light curve models.

Measurement of the angular density of cross-section at the head of the dust cloud (one of the free parameters of the exponential-decay distribution) gives the planet's dust {mass-loss rate} $\dot{M}_{\rm d}$ \citep[see Eq.~(21) of][]{2014A&A...572A..76V}.
Alternatively, $\dot{M}_{\rm d}$ may be estimated directly from the transit depth by adopting physically plausible values for
quantities such as the size, density, and survival time of the dust grains \citep{2012ApJ...752....1R,2014ApJ...784...40R,2013MNRAS.433.2294P,2013ApJ...776L...6K}.

The length of the dust tails (as parametrised by the characteristic angle of decay $\theta_{\rm tail}$) is determined by the sublimation rate of the dust grains \citep[see Eq.~(22) of][]{2014A&A...572A..76V} and radiation pressure parameter $\beta$. Since sublimation rates depend on material properties, the {tail length} can be used to probe the {composition and size of the dust}. The inferred tail lengths of KIC~1255b and \mbox{KOI-2700b} indicate that the dust released by these planets may consist of corundum (Al$_2$O$_3$) or, for \mbox{KOI-2700b}, fayalite (Fe$_2$SiO$_4$). Some other compositions, like pure graphite and iron, are found to be less likely because such grains would sublimate too slowly or rapidly to yield the observed tail length \citep{2014A&A...572A..76V,2016A&A...596A..32V}.

\begin{figure}
\centering
\includegraphics[scale=.35]{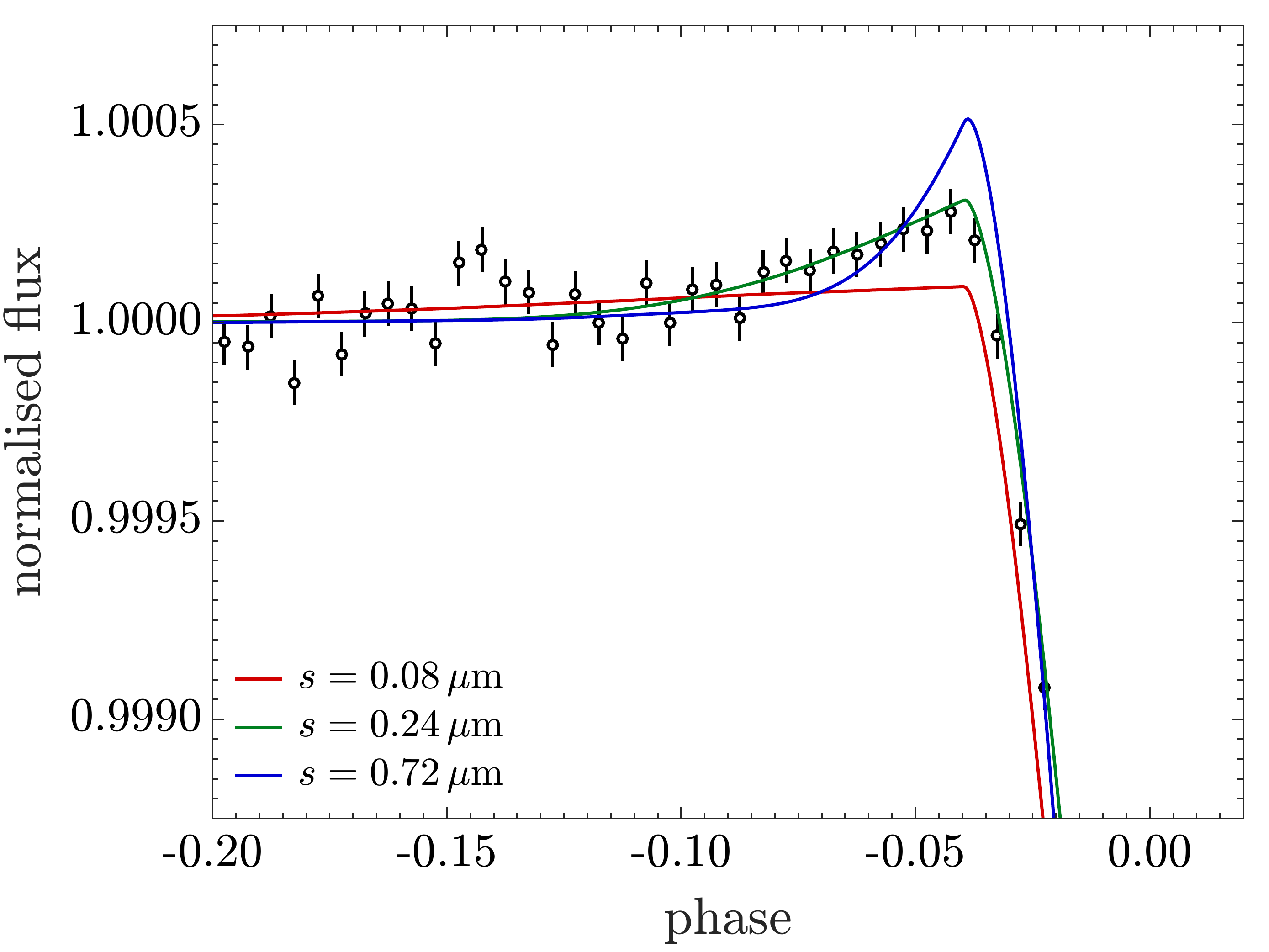}
\caption{Zoom-in of the pre-ingress brightening of KIC~1255b, together with three model light curves using different grain sizes. The variation in the shape of the scattering bump is caused by the dependence of preferred scattering angles on grain size. }
\label{fig:scat}       
\end{figure}

Further constraints on the dust properties come from the detailed shape of the {forward-scattering bumps}, which is sensitive to the grain-size-dependent scattering phase function \citep[e.g.][]{2016MNRAS.461.2453D}.
Larger grains scatter more strongly in the forward direction,
and hence give rise to higher, sharper bumps (see Fig.~\ref{fig:scat}). Exploiting this dependency, \citet{2012A&A...545L...5B} and \citet{2013A&A...557A..72B} derived typical grain sizes in the range 0.1$-$1\,$\upmu$m for KIC~1255b and \citet{2015ApJ...812..112S} find grain sizes of 0.3$-$1\,$\upmu$m for \mbox{K2-22b}.

Dust sizes can also be constrained via the detection of a {wavelength dependence of the transit depth}. The extinction cross-section of dust grains depends on the ratio of their size and the wavelength of the radiation (as well as on the complex index of refraction of the dust material), with larger grains remaining opaque out to longer wavelengths. Therefore, a dust cloud in which the cross-section distribution is dominated by large dust (i.e., larger than the wavelength of the observations) will have transits whose depth is independent of wavelength. For dust clouds in which small dust dominates the cross-section, a colour dependence of the transit depth can be detected. Assuming that the grain-size distribution follows a power law, the slope of the colour dependence can be used to put constraints on the slope of the grain-size distribution and the maximum grain size \citep[see Sect.~5.1 of][]{2014ApJ...786..100C}.

For KIC~1255b, several attempts at measuring a colour dependence of the transit depth have been made. \citet{2014ApJ...786..100C} and \citet{2016ApJ...826..156S} found no significant evidence for a colour dependence, setting lower limits on the typical grain size of $\sim$0.2$-$0.5\,$\upmu$m.
\citet{2015ApJ...800L..21B} find some evidence, during a single observation, for colour dependence which is consistent with a grain size distribution resembling that of our interstellar medium.
For \mbox{K2-22b}, \citet{2015ApJ...812..112S} found a colour dependence during one of three observations which suggests dust with a non-steep size distribution and maximum grain sizes in the range $\sim$0.4$-$0.7\,$\upmu$m.

Finally, we note two caveats regarding the above inferences. Firstly, most models for explaining the shape of the transit profiles have focused on fitting {\em average} transit profiles, while the profiles may possibly vary from transit to transit by more than just a scaling with transit depth.
An exception to this is the work of \citet{2014A&A...561A...3V}, which analysed individual transit events of KIC~1255b. Secondly, the dust cloud is always assumed to be optically thin along the line of sight. For KIC~1255b, there are indications that this does not hold very close to the planet \citep{2016A&A...596A..32V}. {Optical depth effects} may affect the shape of the transit profile, including that of the forward-scattering bumps, as well as the wavelength dependence of the transit depth.

\section{Physics of the Planetary Outflow}

The dust clouds that cause the peculiar transit profiles in the {\em Kepler} data originate in planets that are too small to be detected {\em directly} in the same light curves.  In this section we describe studies of how the dust can escape these planets, and what can be learned about the planets themselves.

As described in the previous section, the dust mass-loss rate $\dot{M}_{\rm d}$ of the planets (i.e., excluding the mass lost as gas) can be derived from their transit profiles. KIC~1255b and \mbox{K2-22b} both have rates of $\dot{M}_{\rm d} \sim 1 $\,M$_{\oplus}$\,Gyr$^{-1}$ in dust alone; \mbox{KOI-2700b} emits dust at a rate one or two orders of magnitude lower. Since the {mass-loss rate} $ \dot{M} $ of a planet depends on its mass $ M_{\rm p} $, comparing these numbers to mass-loss rates that are calculated theoretically, it is possible to put constraints on the mass and lifetimes of these bodies.

\begin{figure}
\centering
\includegraphics[height=7.5cm]{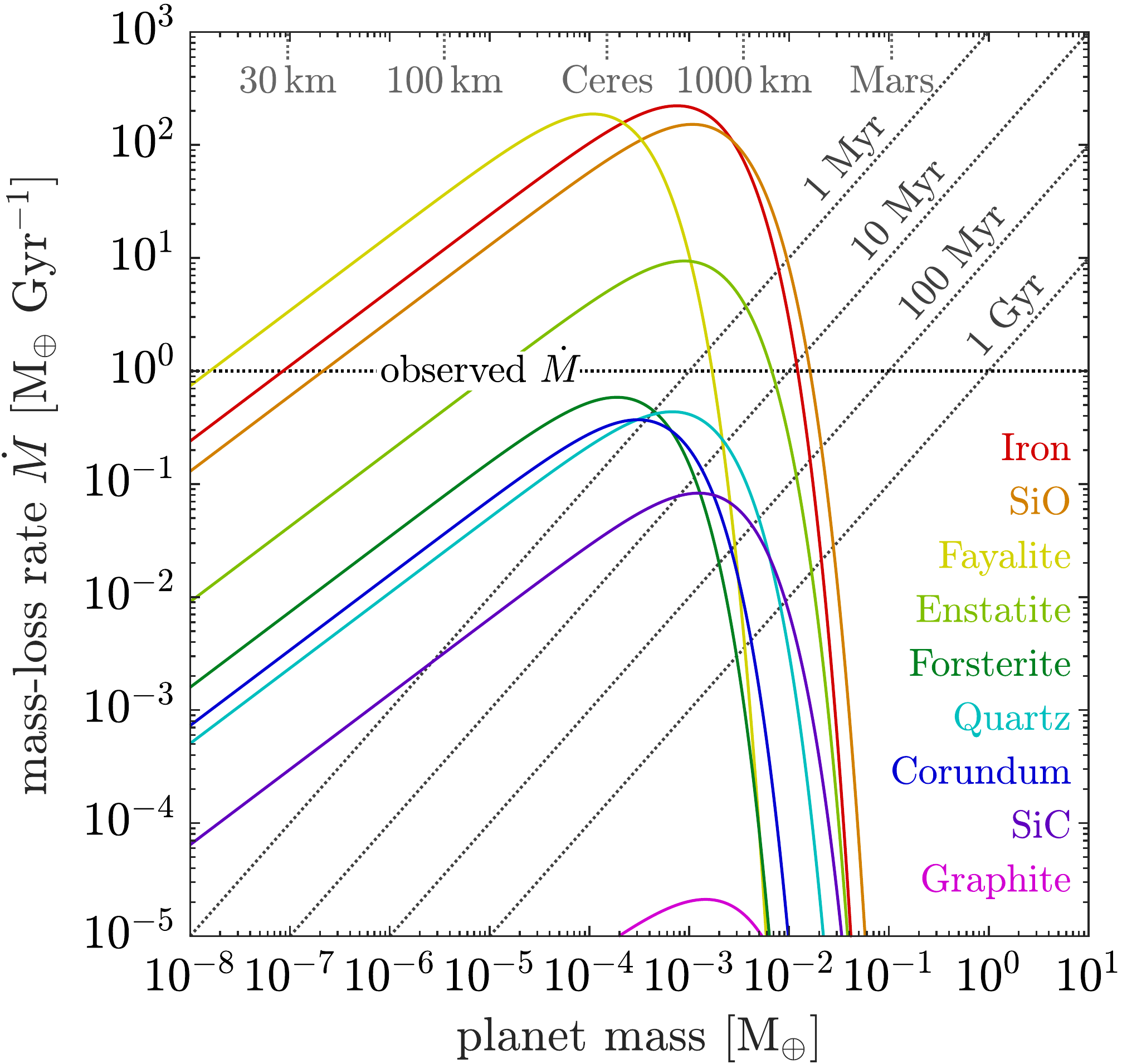} \hglue0.01cm
\includegraphics[height=7.5cm]{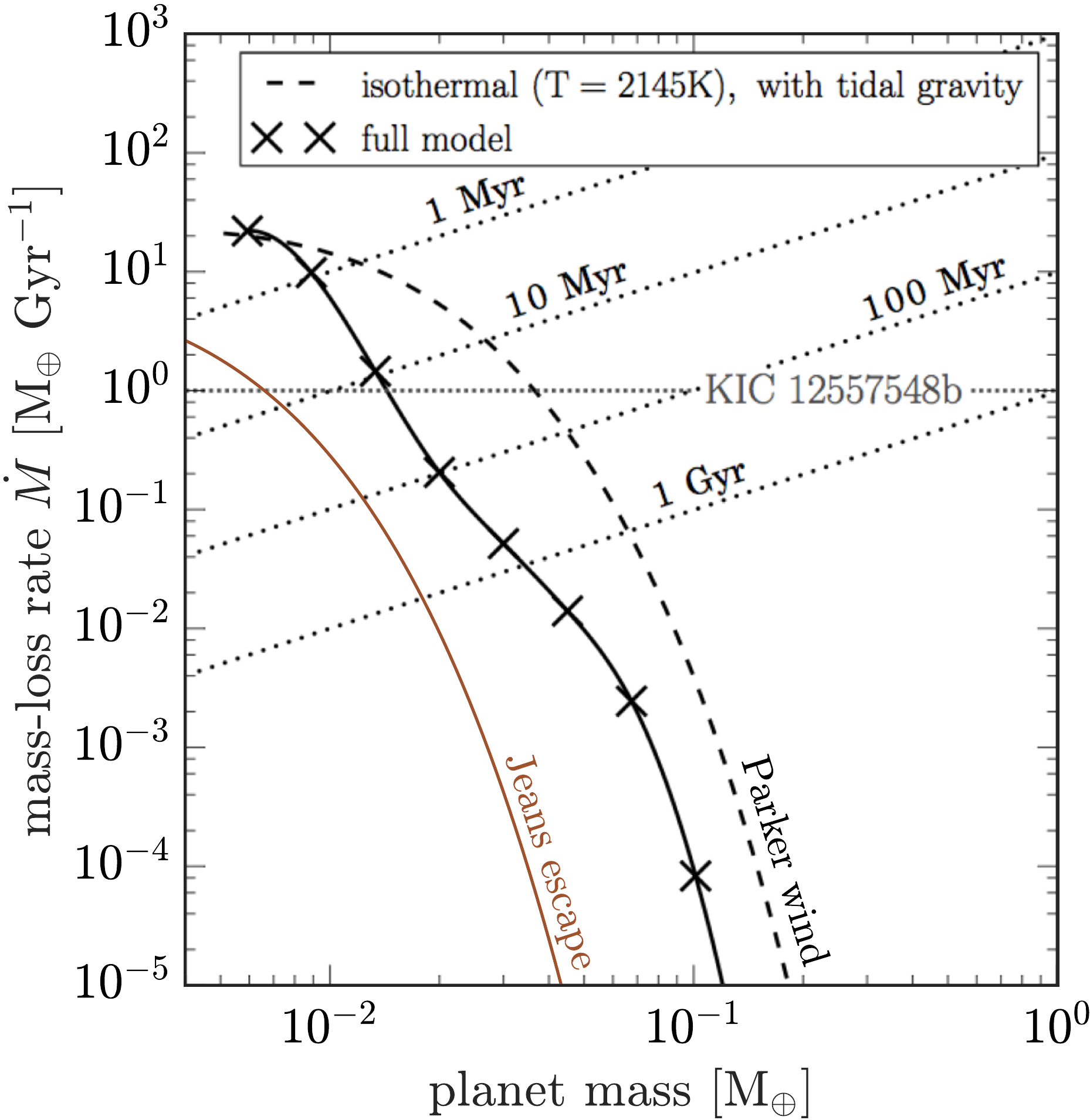}
\caption{Mass-loss rates as a function of planet mass for different mass-loss mechanisms. {\bf Left panel:} The free-streaming and Jeans-escape regimes for different minerals, using an illustrative surface equilibrium temperature of 2100\,K. {\bf Right panel:} Comparison of mass-loss rates due to a Parker wind (black lines) and Jeans escape (brown line), both computed using the same material properties (adapted from Fig.~8 of \citealt{2013MNRAS.433.2294P}). In both panels, the diagonal dotted lines indicate contours of constant disintegration lifetime. The horizontal dotted lines indicate the estimated mass-loss rate of KIC~1255b and \mbox{K2-22b}. }
\label{fig:mdot}       
\end{figure}

At the low-mass end, gas can flow freely from the body, possibly carrying dust along with it from the planet's surface, much like the mass loss from a cometary nucleus. In this {`free-streaming' regime}, the mass-loss rate depends on the surface area of the body available for sublimation, and hence follows an $ \dot{M} \propto R_{\rm p}^2 \propto M_{\rm p}^{2/3} $ power law. The rate also depends strongly on the planet's surface temperature, and on its composition (via the sublimation properties of the material as well as its molecular weight).  A plot of $\dot M$ as a function of the body's mass (and, implicitly, its size) is shown in the left panel of Fig.~\ref{fig:mdot} for a range of different minerals.

This `free-streaming' regime comes to an end quite abruptly (transitioning to the {`Jeans-escape'} regime) when the planet's mass becomes sufficiently high for its self-gravity to play a dominant role. This happens roughly when the planet's escape velocity exceeds the mean thermal velocity of the gas particles.
Typical values for the thermal and escape velocities are:
\begin{align*}
v_{\rm th}
& \simeq 0.8\;{\rm km~s^{-1}}\;
  \bigg( \frac{ T_{\rm eff,p} }{ \rm 2000\;K } \bigg)^{1/2}
  \bigg( \frac{ \mu }{ 50 } \bigg)^{-1/2} \\
v_{\rm esc}
& \simeq 2.4\;{\rm km~s^{-1}}\;
  \bigg( \frac{ \rho_{\rm p} }{ \rm 5\;g/cm^3 } \bigg)^{1/6}
  \bigg( \frac{ M_{\rm p} }{ \rm 0.01\;M_\oplus } \bigg)^{1/3},
\end{align*}
where $T_{\rm eff,p}$ is the planet's equilibrium temperature, $\mu$ mean molecular weight, $\rho_{\rm p}$ the planet's bulk density, and $M_{\rm p}$ planet mass.
For most relevant gas species, this transition to the {Jeans mass-loss} mode happens around planet masses of $ M_{\rm p} \sim 10^{-4}$--$10^{-2} \, $M$_\oplus$, when $v_{\rm esc}$ becomes significantly greater than $v_{\rm th}$ and gravity quenches the mass-loss rate, causing an exponential decrease in $\dot M$ with increasing planet mass.  

The left panel of Fig.~\ref{fig:mdot} shows this exponential termination of the free-streaming region.
We can also see from this set of curves that the maximum distintegration lifetime ($M_{\rm p}/\dot M$), with mass-loss rates of interest occurring for some of the more volatile minerals, is limited to masses of $\lesssim 0.01 \, $M$_\oplus$, and maximum total lifetimes of only $\sim$10\,Myr.  For such short lifetimes (compared to the stellar ages of Gyr), this would imply quite small likelihoods of finding dusty-tailed planets.

Notwithstanding the limitations imposed on mass-loss rates due to Jeans escape, there is another effect that already sets in when $v_{\rm th}$ is only $\sim$$v_{\rm esc}/4$ that is capable of driving large mass outflows.  This mechanism is the {Parker wind} \citep{1958ApJ...128..664P}. It is not an escape of the highest thermal velocity molecules, but rather a {\em hydrodynamic} outflow driven by pressure gradients.  This has been discussed in the context of dusy-tailed planets by \citet{2012ApJ...752....1R} and in much more detail by \citet{2013MNRAS.433.2294P}.  Figure~\ref{fig:mdot} shows an illustrative result from the latter work, together with the Jeans-escape mass-loss rate computed using the same parameter values as used by \citet{2013MNRAS.433.2294P}. The figure demonstrates that, compared to Jeans escape, a Parker wind can produce the observed mass-loss rates for planets with higher masses, and hence longer disintegration lifetimes. This yields a more plausible likelihood of detection over the lifetime of the host star.

Presumably the Parker wind that is driven off the planet is comprised largely of the {metal vapors} emanating from its irradiated, very hot, and likely molten surface.  These have been referred to as `{lava or magma oceans}', and have been modeled by \citet{2011Icar..213....1L} and \citet{2016ApJ...828...80K}.  The compositions of the atmospheres surrounding these lava oceans have been discussed by \citet{2009ApJ...703L.113S}, \citet{2011ApJ...742L..19M}, and \citet{2012ApJ...755...41S}. At the present time it is not clear whether the dust grains are formed in the planet's atmosphere and then driven out entrained in the Parker wind \citep{2012ApJ...752....1R,2013MNRAS.433.2294P}, or rather whether the gas condenses out into grains once the gas has escaped, expanded, and cooled (see Sect.~4.2 of \citealt{2013MNRAS.433.2294P} and Sect.~5.2 of \citealt{2014ApJ...786..100C}).

The observed variable extinction (including the orbit-to-orbit {variations in transit depth} exhibited by KIC~1255b and \mbox{K2-22b}) can naturally be explained by dust as opposed to by hard bodies.  It is easy to envision that dust emission might well not be a steady process, and indeed might be quite erratic.  As one example of a limit-cycle process, consider the following.  When no dust is present, the irradiation by the host star would heat the planet's surface and start the outflow of material.  However, once too much dust has been produced, the extinction of the starlight might quench the heating of the planet's surface, thereby reducing (or stopping) the dust production.  Once the dust clears, the heating and dust-production cycle would start anew \citep{2012ApJ...752....1R,2013MNRAS.433.2294P}.  

Finally, in regard to variable dust emission, \citet{2013ApJ...776L...6K} detected a correlation between the transit depths in KIC~1255b and the rotational phase of the host star (with its 22.9-day period).  They proposed that the host star has regions of enhanced XUV emission, and when these pass closest to the orbiting planet there is a corresponding enhancement of the dust emission.  On the other hand, \citet{2015MNRAS.449.1408C} proposed that this effect may rather be due simply to the passage of the dust tail over large starspot features on the host star.

\vspace{0.2cm}
\noindent

\section{Prospects for Future Research}

In the future, there are at least three ways to further pursue our understanding of this class of disintegrating planets.  The first would be additional ground-based observations of the three known dusty-tailed planets.  The second would be to use future space-based observations to discover more of these objects.  Finally, theoretical investigations into the formation process for such planets could yield additional insight, and suggest further critical observations.

The next observational steps would include better attempts to understand the size distribution and composition of the dust grains.  As we have discussed, measurements of the transit depths vs.~wavelength can, in principle, tell us something about the grain size distribution, and possibly indirectly about the mineral composition of the dust.  The difficulty is that the transit depths are rarely greater than $\sim$1\% and, if one requires fractional depth measurements that are good to a percent, this becomes very challenging to do from the ground.  There is also the possibility that with sufficiently large telescopes it will be possible to measure (i) atomic absorption lines from metals that represent the gaseous part of the same effluents producing the dust, and (ii) polarisation effects due to dust scattering.  The first would more directly inform us about the chemical composition of the molten surfaces of the disintegrating planets, while the latter would provide direct evidence for the dust grains.  Finally, we note that the measurement of a single transit often requires telescope time for a whole night.  And, due to the irregularity of the transits, it makes both planning and requests for telescope time difficult.  

In the near future, observations with TESS \citep{2014SPIE.9143E..20R} seem like a promising way to increase the number of such objects that are known.  TESS should study more stars than were targeted by {\em Kepler} and K2 combined.  The orbital periods of these disintegrating planets are expected to be short (i.e., $\lesssim 1$\,day) and therefore the typical TESS observation interval of one month is more than sufficient to discover these objects.  The added bonus is that the host stars are likely to be brighter and therefore easier to do follow-up studies from the ground.  Finally, we note that there is more phase space available to detect some of these objects even if the planet and its dusty tail do not transit the host star via the `forward scattering' signal alone \citep[see][]{2016MNRAS.461.2453D}.

Formation scenarios for close-in rocky planets are not fully understood.  Possibilities include the formation of a rocky planet much farther out in the system than their current location, and their subsequent disk migration in toward the host star.  More recently, \citet{2017ApJ...842...40L} have suggested that small rocky planets with very short orbital periods (1$-$10\,days) may actually form in situ in the protoplanetary disks and can be subsequently brought in to even shorter periods ($\lesssim 1$\,day) via tidal interactions with the host star.  Additionally, Neptune-like planets that migrate close to the host star could subsequently lose their outer gaseous envelopes via photoevaporation processes \citep[e.g.][]{2013ApJ...776....2L,2015ApJ...813..101V}.  Theoretical investigations along these lines will promote, and in turn be stimulated by, new observational discoveries of these objects.

~

{\small

\noindent \textbf{Acknowledgements}
RvL acknowledges support from the European Union through ERC grant number 279973.

}

\section{Cross References}

{\small

\noindent Barman, T, Planetary Evaporation Through Evolution

\noindent Borucki, W, Space Missions for Exoplanet Science: Kepler / K2

\noindent Vanderburg A \& Rappaport S, Transiting Disintegrating Planetary Debris around WD 1145+017

}


\bibliographystyle{spbasicHBexo}
\bibliography{bib_ads}

\end{document}